# The Global Majority in International AI Governance

Chinasa T. Okolo, Ph.D., Technecultura[1]

Mubarak Raji, School of Information Sciences, University of Illinois Urbana-Champaign[2]

## Abstract

This chapter examines the global governance of artificial intelligence (AI) through the lens of the Global AI Divide, focusing on disparities in AI development, innovation, and regulation. It highlights systemic inequities in education, digital infrastructure, and access to decision-making processes, perpetuating a dependency and exclusion cycle for Global Majority countries. The analysis also explores the dominance of Western nations and corporations in shaping AI governance frameworks, which often sideline the unique priorities and contexts of the Global Majority. Additionally, this chapter identifies emerging countertrends, such as national and regional AI strategies, as potential avenues for fostering equity and inclusivity in global AI governance. The chapter concludes with actionable recommendations to democratize AI governance for Majority World countries, emphasizing the importance of systemic reforms, resource redistribution, and meaningful participation. It calls for collaborative action to ensure AI governance becomes a catalyst for shared prosperity, addressing global disparities rather than deepening them.

---

[1] Chinasa T. Okolo is the Founder and Scientific Director of Technecultura, Washington, D.C. https://orcid.org/0000-0002-6474-3378

[2] Mubarak Raji is a Nigerian licensed attorney and PhD student at the School of Information Sciences, University of Illinois at Urbana-Champaign, United States. https://orcid.org/0009-0003-0287-4393

# 1. Introduction

The proliferation of artificial intelligence (AI) technologies is profoundly reconfiguring relationships between the Global Majority and Western countries, often exacerbating entrenched asymmetries of power (Veale et al., 2023; Okolo, 2023a; Peng, 2024). In this chapter, we classify the "Global Majority" as countries within Africa, Asia, the Caribbean, Latin America, and Oceania that have experienced socioeconomic marginalization and where communities have been racialized into ethnic minority groupings. Much like the Industrial Revolution, which entrenched economic disparities, AI risks creating even more pervasive and accelerated forms of inequality. The development of AI demands immense computational resources, vast and diverse datasets, and advanced technical expertise, assets disproportionately concentrated in the West, particularly within countries such as the United States, Canada, and the United Kingdom, with China as a notable exception (Okolo, 2023b). This uneven distribution engenders a systemic technological dependency, wherein nations in the Global Majority are compelled to adopt AI systems designed and optimized for contexts far removed from their respective socio-cultural and economic realities (Anthony et al., 2024). This dependency manifests starkly across multiple sectors. When Global Majority countries seek to implement AI solutions, they often rely on systems trained on datasets reflective of Western contexts (Prabhakaran et al., 2022). Such systems frequently fail to account for region-specific conditions, infrastructures, or culturally embedded practices, leading to suboptimal outcomes. This dynamic perpetuates a contemporary form of technological colonialism, wherein Global Majority countries remain subservient to external technological paradigms. The economic implications of this disparity are equally profound. As AI assumes a pivotal role in driving global economic competitiveness, the chasm between AI-enabled and AI-dependent nations widens. Corporations headquartered in the West exploit their technological advantages to dominate markets in the Global Majority, often to the detriment of local innovation ecosystems and economic self-determination (Kwet, 2021; Hadgu et al., 2023).

The governance of AI at the global level reflects a pronounced centralization of authority among a limited set of actors. Dominant technology corporations, primarily based in the United States and China, exercise outsized influence by controlling critical AI infrastructure and spearheading cutting-edge research. Companies such as Google, Microsoft, OpenAI, and Baidu set de facto standards and development trajectories that subordinate other nations and entities to their frameworks. Simultaneously, Western countries, notably the United States and those within the European Union, lead in shaping AI regulatory paradigms and governance architectures (Leslie et al., 2024). The European Union's AI Act exemplifies this dominance, with its global ramifications often described as the "Brussels Effect," whereby EU regulatory standards diffuse internationally (Bradford, 2019). This dynamic compels Global Majority countries to conform to frameworks that often overlook their unique challenges and priorities. The disproportionate allocation of AI investment in the countries at the forefront of AI development (also referred to as leading AI nations or frontier AI powers) creates a reinforcing cycle where economic and governance power are mutually reinforcing. States and corporations with superior resources not only advance their technological edge but also cement their dominance over governance structures, perpetuating asymmetries of influence. Furthermore, the preeminence of tech companies and research institutions in frontier AI powers grants these countries a dominant voice in determining the trajectories of AI development, including ethical principles, technical standards, and governance norms. This intellectual dominance marginalizes alternative epistemologies and priorities that may better reflect the needs and values of Global Majority contexts.

The centralization of power in AI governance raises profound questions about equity and justice in global development. However, emergent countervailing trends suggest opportunities for a more balanced global AI ecosystem. Aside from China, Global Majority countries like India are cultivating significant AI capabilities, thereby challenging the unipolar dominance of traditional AI powerhouses (Chahal et al., 2021). Regional bodies, such as the African Union and ASEAN, are increasingly articulating indigenous AI strategies and governance frameworks, signaling a nascent shift toward more localized and inclusive paradigms (African Union, 2024; ASEAN, 2024). Moreover, a growing consensus from civil society researchers underscores the necessity for equitable governance models that prioritize diverse perspectives and shared benefits (Okolo et al., 2023a; Adan et al., 2024; LaForge et al., 2024). A rising number of grassroots initiatives and multilateral collaborations within Global Majority countries aimed at democratizing AI also hold promise for mitigating these disparities (Mbayo, 2020). This chapter examines critical questions that address the disparities and power dynamics inherent in global AI governance. What are the existing disparities in AI development that impact Global Majority countries, and how do they shape governance structures on a global scale? How can governance frameworks be designed to integrate and prioritize diverse perspectives across the Global Majority, ensuring their unique contexts and priorities are adequately represented? Furthermore, what institutional mechanisms could be developed to amplify the agency of Global Majority countries in influencing AI development and governance?

In Section 2, this chapter discusses inequities in education, research capacity, and infrastructure that disproportionately impact the ability of Global Majority countries to contribute to AI development. Section 3 details efforts by Global Majority countries to draft governance frameworks and enact AI legislation. Section 4 then covers the landscape of global AI governance efforts, including international conventions, treaties, declarations, standards bodies, and high-level convenings, examining the participation of Global Majority countries in these respective initiatives. Section 5 describes efforts led by Global Majority countries to contribute towards more inclusive AI governance, outlining key events and stakeholders within this domain. Section 6 synthesizes the chapter by providing recommendations for governments to increase capacity-building efforts to strengthen AI research and advance AI policy development. Overall, this chapter aims to chart pathways toward a more equitable and inclusive framework for global AI governance by addressing pivotal questions around inequality in AI research and development. In doing so, it also seeks to highlight actionable strategies to mitigate existing inequalities and forestall the exacerbation of global divides.

## 2. Inequality in Global AI Development

The rapid adoption of artificial intelligence has exposed the impact of persisting development challenges in education, poverty, and infrastructure development, turning what we know as the digital divide into the AI divide. This section details systemic challenges that impact AI research, development, and innovation in Global Majority countries, detailing disparities in access to AI education, AI research production, and digital infrastructure costs. Understanding these inequalities is crucial for developing effective strategies to make AI development and, furthermore, AI governance more inclusive and equitable.

### 2.1 Educational Challenges in Global Majority Contexts

The stark disparities in AI R&D output reflect and reinforce existing global inequalities. The foundation of this challenge begins with education systems in Low and Middle-Income Countries (LMICs), where limited resources and infrastructure often result in inadequate STEM education. In Low and Middle-Income countries, many of which are in the Global Majority countries, 70% of children are estimated to experience learning poverty —a metric detailing a child's inability to read and comprehend simple texts by 10 years of age (Saavedra et al., 2022). Disparities in access to quality education are also persistent regarding tertiary education, such as Bachelor's, Master's, and PhD-level degree programs. Many top universities offering AI-related degree programs are in the United States and the United Kingdom (QS World University Rankings, 2024), with many of these courses being offered in English (Maslej et al., 2024). The concentration of AI education programs in English-speaking countries creates a significant barrier for many potential AI researchers and developers, effectively excluding a significant portion of the world's population from equitably participating in AI development. While educational institutions within Global Majority countries have increased their respective capacities to offer AI-related courses, many of these universities also lack advanced computing resources, updated curricula, and specialized faculty needed for cutting-edge AI education.

## 2.2 Global Disparities in AI Research and Development

The geographic distribution of AI talent reveals another dimension of inequality in AI R&D. When talented individuals from Global Majority countries relocate to major tech hubs outside their home countries, this migration contributes to brain drain that further weakens local AI development capacity. Countries like China, the United Kingdom, Germany, France, and Canada hold significant shares of the global AI workforce, and it is estimated that 42% of top-tier AI research migrates abroad (MacroPolo, 2023). The United States remains the global hub for AI talent, attracting the largest share of elite AI researchers while also housing top AI companies and research institutions. The concentration of AI talent in a few geographic areas creates a clustering effect, excluding participation from Global Majority countries. Major tech companies in higher-income countries can offer higher salaries and better resources, making it difficult for organizations in Global Majority countries to retain local talent or attract international expertise. However, there has been marginal progress in the distribution of AI talent, with research published in the Harvard Business Review showing that cities such as Bangalore (India), Kuala Lumpur (Malaysia), Johannesburg (South Africa), São Paulo (Brazil), Bangkok (Thailand), Jakarta (Indonesia), and Lagos (Nigeria) are emerging as global AI hubs (Chakravorti, 2021).

This educational and talent gap described previously directly impacts the ability of Global Majority countries to meaningfully contribute to AI research production. Countries like China have developed strong AI expertise, resulting in a significant share of AI research articles published in journals, conferences, and online repositories (Maslej et al., 2024). However, most Global Majority countries remain underrepresented at premier research conferences in AI and ML (Okolo, 2023a). The development of prominent AI models further illustrates this divide. Well-resourced companies in the Global North predominantly create models like GPT-4, DALL-E, and Claude. These models, trained on data that often underrepresents Global Majority perspectives, become de facto standards that other countries must use, regardless of their local needs or contexts. AI research and development are also increasingly concentrated in large companies, with 55% of all notable AI models produced from industry (Maslej et al., 2025). The patent landscape tells a similar story, with China and the United States holding the vast majority of AI-related patents (69.7% and 14.2%, respectively), creating significant barriers to innovation in other regions worldwide. Over the past decade, AI patents have grown exponentially, with 122,511 patents granted in 2023 alone (Maslej et al., 2025). This concentration creates a self-reinforcing cycle - these companies attract more talent, secure more funding, and further cement their leadership positions in the global AI race.

## 2.3 Compute and Digital Infrastructure Divide

The AI divide becomes particularly acute when considering the infrastructure requirements for AI development. Given that the costs to train prominent AI models like Gemini, ChatGPT, and DALL-E are in the millions of dollars and are expected to reach 1 billion dollars by 2027 (Cottier et al., 2024), these exorbitant costs are unreachable for a significant number of research institutions, both within and outside of the Global Majority. Many countries in the Global Majority struggle with basic internet connectivity, with 27% of people living in low-income countries having access to the Internet (ITU, 2024). Despite recent improvements in submarine cable installations, regions like Africa still face significant connectivity challenges, having an internet penetration rate of 37% compared to the global average of 67% (ITU, 2024). This limited connectivity affects everything from data collection to model training to AI deployment. The distribution of data centers presents another critical barrier. The United States has the highest concentration of data centers worldwide, equaling almost half of all data centers worldwide (Statista, 2025). In addition to a lack of AI talent and limited resources to devote to AI development, challenges with access to telecommunications and computing infrastructure make it harder for researchers and developers to collect and process local data, leading to AI systems that underrepresent Global Majority values, cultures, and languages (Bengio et al., 2024; Ghosh et al., 2024; Tao et al., 2024). Overall, these issues of AI R&D concentration create unfair geographic advantages, leading to cascading effects that negatively impact equitable Global Majority participation in AI governance.

## 3. Global Majority AI Governance Efforts

To ensure proper AI oversight, numerous Global Majority countries and regional organizations have developed initiatives on AI governance. The Brussels Effect has increased the global acceptance of EU GDPR as a model data protection framework (Bradford, 2019), also creating a similar impact for the EU AI Act. However, critics of the Brussels Effect on economically developing countries that are part of the Global Majority have raised concerns about distinct societal norms, economic landscapes, and enforcement capacity, which would lead to regulatory misalignments, cultural and contextual mismatch (Mannion, 2020; Canaan, 2023; Bhatt & Burman, 2024; Ferracane et al., 2025). Similar to data protection frameworks, Global North-focused AI governance frameworks, such as the EU AI Act, Global Partnership on Artificial Intelligence (GPAI) Principles, and OECD AI Principles, are gaining momentum without the inclusion of most Global Majority countries (Adan, 2023). Such exclusion could amplify the existing global digital divide and inequalities (Hassan, 2022) and may lead to potential bias, which makes global inclusion a necessity to ensure AI governance does not face the same challenges as data protection (Wicker, 2024). This section describes AI governance initiatives in Africa, Asia, Latin America, the Caribbean, and Oceania, and compares them with the EU Act and OECD AI principles.

### 3.1 Africa

The development of a standalone AI governance legislative framework in Africa is still in the early stages. The African Union has introduced an AI strategy as a regional international framework for AI governance. At the national level, about 13 African countries have introduced AI strategies or draft AI policies through soft laws to boost innovation, development, and a step towards regulating AI (Okolo, 2024b).[3]

In July 2024, the African Union (AU), a continental body of 55 African countries, released the AU Continental Artificial Intelligence Strategy to foster social and economic development after it was endorsed by its Executive Council in Ghana (African Union, 2024). The AU-AI Strategy

---

[3] As of July 2025, these countries are Algeria, Benin, Egypt, Ethiopia, Ghana, Kenya, Lesotho, Mauritius, Nigeria, Rwanda, Senegal, Tunisia, and Zambia

covers five core areas: increasing the benefits of AI for African development in public and private sectors; strengthening Africa's capabilities to boast technological infrastructure, literacy, AI expertise and skills, and research; reducing AI risk through AI governance while adhering inclusivity, human rights, African cultural values, for the deployment of ethical, safe and secure AI; government and industry collaboration; and regional and international collaboration (African Union, 2024). The AU-AI Strategy was informed by dozens of experts across the continent, and it was developed specifically to tackle Africa's distinctive socio-economic barriers with African-driven solutions and enhance local talents. It prioritizes recognizing African values, such as the concept of Ubuntu, and emphasizes people-centric principles in building a robust AI ecosystem (African Union, 2024).

While acknowledging the need to meet international best practices on AI governance, the AU-AI Strategy categorized risks based on the potential harms they can cause African societies. It highlighted four significant risks that require a multi-tiered approach to address: environmental, system-level, structural risks, and risks to African values (African Union, 2024). While the AU-AI Strategy shares some linguistic features with the EU AI Act (Silva et al., 2025), its risk categorization differs from the EU risk classification. Africa has experienced several environmental pollutions from mining, e.g., Congo (Davey, 2023), and oil spillage, e.g., Nigeria, which is championed primarily by Global North-based companies (Laville, 2024). Hence, addressing the environmental risks of AI systems, such as e-waste, carbon emissions, and the possibility of data centers threatening water supply, is crucial for Africa (African Union, 2024). This indeed reflects the strategy's aim to address Africa's peculiar challenges.

System-level risks address data protection issues, bias, and discrimination in deploying AI. On the other hand, structural risks seek to tackle unemployment, intellectual property, gender equality, and the AI access gap. On risks to African values, the strategy emphasizes the need to tackle the misuse of AI to spread misinformation, efforts that undermine human rights and democracy, and the erosion of African native knowledge and cultural identity. Overall, the AU-AI Strategy follows an African-centric approach to building capacity, fostering development, and ensuring the ethical deployment of AI, as against the EU risk-based approach or OECD ethical guidelines. Additionally, the AU-AI Strategy is best described as non-binding guidance that lays the foundation for a binding regional instrument.

Furthermore, Nigeria and Kenya have bills on AI regulations pending before their legislative houses. On 3 December 2024, four consolidated AI bills passed the second reading stage at the Nigerian House of Representatives (House of Representatives, Nigeria, 2024).[4] More particularly, the Control of Usage of Artificial Intelligence Technology in Nigeria Bill, 2023 (HB.942), seeks to ensure the ethical deployment of AI in Nigeria to enhance societal safety. Section 3 of the bill enumerates ethical principles for AI system must prioritize and advance to human values, human rights, transparency, and fairness; adhered to human privacy, safety, and data protection laws; prohibit discrimination; accountability of AI users and developers for operational consequences. Unlike the OECD AI principles, the bill does not have provisions on sustainability and inclusive growth, which creates a vacuum on how it will address AI inequalities and its likely impact on the environment.

The bill also provides for risk assessment and mitigation, educational awareness, compliance with data protection and data governance principles, safety and security, and penalties. Although the bill envisages risk assessment and mitigation in AI systems, it fails to categorize risks into different levels, like the EU Act, which expressly categorizes risks into unacceptable, high, limited, and minimal risks. In summary, the bill lacks comprehensive legal requirements and technical specifications to address the fundamental aspects of developing and deploying AI systems that align with ethical standards.

The Kenya Robotics and Artificial Intelligence Society Bill of 2023 seeks to establish a society for advancing and deploying ethical AI and robotics in Kenya and provides licensing requirements for these technologies. The bill highlights five principles of "public good," "human safety and security," "privacy and data protection," "accountability," and "diversity and inclusion" (Robotics Society of Kenya, 2023). However, the bill has received positive appraisals and criticism, such as its narrow regulatory approach and lack of effective enforcement mechanisms, which may hinder innovation (Sang, 2024; Kangwana & Tuitoek, 2024).

To address bias, power imbalances, and reduce the influence of Global North approaches to AI governance in Africa and by extension the Global Majority, a number of scholars have advocated for a decolonial approach to AI governance (Coleman, 2023; Hassan, 2022). Rwanda adopted this approach in its national AI strategy, and there is a likelihood that more countries will follow the pattern (Ayana et al., 2024; Silva et al., 2025). Additionally, about 38 out of 55 African countries have data protection laws with provisions that regulate personal data processed through automated means (ALT Advisory, 2022; Okolo, 2024a; Raji et al., 2025). However, these data protection laws are limited to protecting personal data privacy and are insufficient to regulate AI. Data privacy is essential in bolstering comprehensive AI regulation and should receive equal attention from countries within Africa and across the Global Majority.

---

[4] The bills are the Establishment National Institute for Artificial Intelligence and Robotic Studies Sciences Regulation Commission, Control of Usage of Artificial Intelligence Technology in Nigeria, Regulate the Development, Deployment, and use of Artificial Intelligence in Nigeria and for Related Matters (HBs.143, 601, 942, and 1810 respectively) sponsored by Honorables Sada Soli, Benjamin O. Kalu, Ademorin A. Kuye, and Ibrahim Ayokunle Isiaka.

## 3.2 Asia

Asia has significant concentrations of AI talent and produces large amounts of AI research, primarily from countries such as China, Singapore, Japan, India, and Taiwan. There are currently no continent-wide AI governance initiatives, like the AU-AI Continental Strategy, that are targeted towards the Asian continent. However, regional efforts, such as the Association of Southeast Asian Nations Guide on AI Governance and Ethics, have begun to emerge. Several countries, such as Bangladesh, China, India, Indonesia, Pakistan, Taiwan, Thailand, and Vietnam, have introduced national AI strategies or plans (IAPP, 2024; Keith, 2024).

The Association of Southeast Asian Nations (ASEAN) is a multilateral regional bloc for economic integration and regional development in Southeast Asia and comprises ten member countries.[5] ASEAN released its Guide on AI Governance and Ethics in February 2024 as a regional AI governance framework for adoption by organizations and national governments in the development and deployment of AI in member countries (ASEAN, 2024; Keith, 2024; Putra, 2024; Xu et al., 2024). The guide set seven guiding principles similar to OECD AI principles: transparency and explainability, fairness and equity, security and safety, robustness and reliability, human-centricity, privacy and data governance, and accountability and integrity (ASEAN, 2024). Similarly to the AU-AI Strategy, the ASEAN guide also focuses on capacity-building to enhance AI expertise and reskill the workforce through continuous education, private sector collaboration, and improving school curriculum, encourage investing in AI companies and support AI innovation networks, increase public awareness of the societal impact of AI, encourage funding of AI research and development, and encourage the implementation of the guide, especially by businesses (ASEAN, 2024).

---

[5] ASEAN member countries are Brunei, Cambodia, Indonesia, Laos, Malaysia, Myanmar, the Philippines, Singapore, Thailand, and Vietnam.

Like the EU AI Act, ASEAN adopted a risk-based approach, which requires human oversight of the AI systems, especially when high risk is involved, while focusing on severity and probability of harm. AI systems risks are divided into three categories: a) minimal risk, which allows the AI system to perform independently with very little human oversight; b) medium risk, which requires necessary human intervention based on the sector or kind of AI systems; and c) high risk requires human control at all times to avert harmful decisions. Unlike the EU AI Act, the Guide does address unacceptable risks that may pose a future challenge or misuse. One significant distinction from the AU-AI Strategy compared with the ASEAN Guide is that it provides practical guidance with different case studies and AI risk impact assessment templates, which can be easily adapted by organizations to conduct risk management. However, economic and development disparities, democratic values, and AI readiness among member countries may hinder the implementation of the guide (Keith, 2024; Putra, 2024).

Few countries within Asia have draft AI legislation in circulation, such as the Artificial Intelligence Law of the People's Republic of China introduced by law scholars, the proposed Digital India Act, 2023, which aims to replace the outdated Information Technology Act of 2000 while encouraging comprehensive oversight of emerging technologies, and Taiwan's AI Basic Act which aims to establish a comprehensive AI governance framework, promoting sustainable development, transparency, and innovation while addressing legal challenges and balancing human rights and risk management (Ho et al., 2024; IAPP, 2024). Other countries are also making regulatory efforts and have working drafts, such as Malaysia's Artificial Intelligence Governance and Ethics Guidelines; Thailand's drafts of the Promotion and Support for Artificial Intelligence Act and the Royal Decree on Business Operations that use Artificial Intelligence Systems; and Vietnam's Digital Technology Industry Law (Ho et al., 2024).

According to the 2025 Stanford AI Index, China is a leader in AI research publications, patents, and industrial robot installations (Maslej et al., 2025). In 2024, China published the AI Safety Governance Framework to enhance international collaborations as part of the Global AI Governance Initiative (China National Technical Committee 260 on Cybersecurity, 2024). The AI Safety Governance Framework outlines AI safety principles that prioritize innovation, creating governance frameworks, and "safe, reliable, equitable, and transparent AI" for human benefit and classification of risks (AI Safety Governance Framework, 2024, pg 2). China has also introduced case-specific AI governance frameworks such as the Algorithmic Recommendation Management Provisions and Interim Measures for the Management of Generative AI Services (IAPP, 2024).

The draft Taiwan AI Basic Act, 2024, which was the National Science and Technology Council, takes a risk-based approach to AI regulation like the EU AI Act, puts innovation first, and saddles the government with the responsibility of creating an accountability structure (Tseng, 2024). To enhance innovation, the draft law permits regulatory testbeds and sandbox innovators (Tseng, 2024). This effort could be connected to Taiwan's role in the global supply of semiconductors, which is integral to developing AI systems. The draft law also addresses the issues of harm, human safety, data privacy, particularly data minimization, intellectual property, developers' and users' insurance and liability, AI literacy, and education (Tseng, 2024).

## 3.3 Latin America and the Caribbean

The emergence of the Santiago Declaration on AI Ethics in October 2023 initiated regional AI governance initiatives in Latin America, similar to those launched within Africa, Europe, and Southeast Asia. Unlike the AU-AI Strategy, which acknowledges several existing AI governance frameworks worldwide, the Santiago Declaration seeks to align with the UNESCO Recommendation on the Ethics of AI, which explains why the declaration emphasizes a human rights approach to AI governance and adopts all ten UNESCO AI principles. Like the AU-AI Strategy, the declaration recognizes AI as a catalyst for transformative development in the region and hence prioritizes regional needs and values. It was introduced to foster regional dialogue and cooperation on ethical AI deployment, addressing the peculiarities of the region while respecting human rights, data protection, and democratic values, fostering sustainability and inclusivity, avoiding discrimination, addressing inequalities, and promoting innovation (Santiago Declaration, 2023; Maffioli, 2023). It also emphasizes the preservation of indigenous knowledge and cultural heritage in the age of AI, mirroring the AU-AI Strategy. Although the declaration recognizes the importance of risk mitigation, it does not include provisions on categorizing risks like the EU AI Act, ASEAN guide, and AU-AI Strategy.

Furthermore, countries such as Brazil, Chile, Colombia, Peru, Mexico, and Uruguay have developed AI strategies or policies in the form of soft law (Cisneros, 2024; IAPP, 2024).

Argentina also introduced the Recommendations for Reliable Artificial Intelligence, Provision 2/2023 (IAPP, 2024; Mosqueira & Gonzalez, 2024).

Additionally, a few Latin American countries have drafted pending AI bills before their legislature. The Argentine AI bill followed a risk-based approach like the EU AI Act. Others include Brazilian Bill 2338/2023 to regulate AI, Chile's AI Bill, the Mexican draft AI bill, and Peru's Law 3814 on AI (IAPP, 2024; Mosqueira & Gonzalez, 2024). Brazil's proposed AI law adopts a risk-based approach, categorizes the AI system's risk into excessive risk and high risk, requires preliminary and risk assessment before an AI system is deployed in the market, and prescribes enforcement mechanisms and penalties for violations by AI National Regulation and Governance System (Freitas & Giacchetta, 2024; Mosqueira & Gonzalez, 2024). The Chilean Bill to regulate AI systems, robotics, and related technologies (Bill 15869-19) applies to all operating AI systems in Chile, whether deployed domestically or in a foreign country, and prescribes a risk-based approach and classifies risks into unacceptable, high, limited, and no risks (Mosqueira & Gonzalez, 2024). The Mexico AI bill was introduced in April 2024, which has an extraterritorial reach like Chile's AI bill; it also followed a risk-based approach and classified risk into unacceptable, high, and low risks; intellectual property holders' consents are required before such IP could be used in training AI models, and regulatory approval is required before deployment of AI system (Mosqueira & Gonzalez, 2024).

## 3.4 Oceania

Within the Global Majority, member countries in Oceania remain underrepresented globally in AI governance efforts. Within this section, we define Global Majority countries within Oceania as the Cook Islands, Federated States of Micronesia, Fiji, Kiribati, Nauru, Niue, Palau, Papua New Guinea, Republic of the Marshall Islands, Samoa, Solomon Islands, Tonga, Tuvalu, and Vanuatu. Given the unique development challenges of countries within this region, exacerbated by a reliance on tourism, geographical isolation, and insufficient infrastructure, a significant number of efforts are primarily directed towards broader digital and Information and Communication Technology (ICT) initiatives rather than specific AI governance frameworks.

In 2025, Fiji published its National Digital Strategy, which aims to support the adoption and integration of digital technologies to enhance key sectors like tourism and education and boost their respective economy (Government of Fiji, 2025). While the strategy doesn't mention specific measures to move towards robust AI governance, the document details the government's plan to develop a National AI Framework by 2027, which is likely to include more provisions on how Fiji will govern AI. Papua New Guinea is actively developing a National AI Adoption Framework to guide the responsible, ethical, and inclusive use of AI across the country. In April 2025, the Department of Information and Communications Technology (DICT) announced its efforts to finalize this framework, which will establish policy and governance structures for AI implementation (Kia, 2025). This framework aims to leverage AI for national development while mitigating potential risks like misuse and deepening inequalities.

Given the population and physical size of Global Majority countries in Oceania, intraregional cooperation on AI governance can help these nations increase targeted efforts towards developing national AI strategies and implementing AI regulation. Efforts such as the Lagatoi Declaration, a landmark commitment by Pacific Island Countries and Territories to accelerate digital transformation and leverage information and communications technologies for socio-economic development and regional integration, exemplify a path for future collaboration in AI-focused efforts (Pacific Ministers for Information and Communications Technologies, 2023). Its corresponding action plan, the Pacific ICT and Digital Transformation Action Plan 2024-2030, outlines specific priorities and strategies to achieve these goals across the region and is expected to be endorsed in August 2025. Although countries in Oceania do not have a standalone AI governance framework yet, available national digital strategies and the Lagatoi Declaration indicate that countries in the region are paying attention to data privacy, ethical and responsible deployment of AI, accountability, inclusivity, regional cooperation, and bias, like many other AI governance frameworks across the world. We urge policymakers to consider Oceania's unique and indigenous peculiarities in framing their AI governance framework, integrate human-centric approaches like ASEAN, and prioritize sustainable development like the AU.

## 4. Global Majority Inclusion in Global AI Governance Efforts

Countries such as the United Kingdom and the United States tend to dominate discourse within the global AI governance landscape. More recently, the United Kingdom, South Korea, and France have led high-level convenings on steering the responsible development and governance of AI. While these initiatives have led to notable progress towards establishing AI Safety Institutes and increased multilateral cooperation through several declarations and international agreements, there is a glaring absence of Global Majority leadership. However, with the Indian government set to host the next global AI Summit along with Brazil and South Africa forming AI initiatives within their respective leadership in the G20, there are positive signals towards more inclusive global AI governance. To understand trends in global AI governance, this section examines the participation of Global Majority countries in international conventions, treaties, declarations, standards bodies, and high-level convenings.

## 4.1 AI Regulatory Efforts

### 4.1.1. 2024 Council of Europe Framework Convention on Artificial Intelligence and Human Rights, Democracy and the Rule of Law

The Council of Europe (CoE) Framework Convention on Artificial Intelligence and Human Rights, Democracy and the Rule of Law is the pioneering binding international instrument on AI governance. The Framework was adopted in May 2024, opened for signature in September 2024, and has ultimately been signed by 16 countries. Although no Global Majority countries are signatories to the Convention as of July 2025, Argentina, Costa Rica, Mexico, Peru, and Uruguay participated in the negotiation and drafting of the Convention (Council of Europe, 2024a; Council of Europe, 2024b). Observer and non-CoE member countries may sign this Convention just like eight non-European countries signed the Convention for the Protection of Individuals with regard to Automatic Processing of Personal Data (Council of Europe, n.d.).[6]

## 4.2 AI Standards Bodies

### 4.2.1. 2018 Public Voice's Universal Guidelines for Artificial Intelligence (UGAI)

At the International Data Protection and Privacy Commissioners Conference in 2018, technology policy experts and civil society organizations endorsed the Universal Guidelines for Artificial Intelligence sponsored by the Public Voice Coalition (Boston Global Forum, 2018; CAIDP, 2023). UGAI highlighted twelve principles to serve as guidance for policymakers around the globe.[7] Technology policy experts and civil societies from Global Majority countries such as Afghanistan, Argentina, Bangladesh, Brazil, Chile, Colombia, Comoros, Ecuador, Fiji, Ghana,

---

[6] See Articles 6, 7, 8, 9, 10, 11, 12, and 13 of the Convention for its AI principles.
[7] https://www.caidp.org/universal-guidelines-for-ai/

Hong Kong, India, Kenya, Malaysia, Mexico, Nepal, Nigeria, Panama, Saudi Arabia, and South Africa were in attendance (CAIDP, 2023).

### 4.2.2. 2019 OECD Recommendation of the Council on Artificial Intelligence

The Organisation for Economic Co-operation and Development (OECD) has developed the foremost multilateral principles on AI governance, the Recommendation of the Council on Artificial Intelligence in 2019. The OECD AI Principles aim to promote the adoption of safe, responsible, and trustworthy AI, building upon prior efforts from the OECD to develop privacy and corporate governance principles (OECD, 2019). Chile, Colombia, Costa Rica, and Mexico are OECD members from the Global Majority. Some Global Majority countries are non-OECD members but adhere to the OECD AI principles, including Argentina, Brazil, Egypt, Peru, and Uruguay (OECD, 2019).

### 4.2.3. UNESCO Recommendation on the Ethics of AI

The United Nations Educational, Scientific and Cultural Organization (UNESCO) is committed to promoting shared values through the advancement of culture, education, science, and technology (UNESCO, 2024). The UNESCO Recommendation on the Ethics of AI, adopted in 2021, provides a global framework for ethical AI development and use, emphasizing human rights, dignity, and the promotion of AI for the benefit of humanity, society, and the environment. It has 10 major principles following the human rights perspective, many of which may likely influence many Global Majority countries as they are among UNESCO's 194 member states, and several of them have already adopted UNESCO AI principles (IAPP, 2024; UNESCO, 2024). Such international cooperation amongst Global Majority countries can also be illustrated by the express alignment of Latin America and the Caribbean countries to the UNESCO AI principles in the Santiago Declaration, as previously detailed in Section 3.3.

## 5. Efforts Towards More Equitable AI Governance

To ensure that global AI governance recognizes the nuances that AI technologies bring to diverse societies and how these technologies impact communities across the globe in different ways, Global Majority countries must be fairly represented and meaningfully involved in developing frameworks, agreements, and policies on AI governance. Such inclusion should not revolve around Global Majority countries being invited to participate on the sidelines of international meetings or sign onto agreements after they have been drafted, but playing a significant role in the origination of these efforts. To understand what progress is being made toward this goal, Section 5 describes efforts led by Global Majority countries to contribute towards more inclusive AI governance, outlining pivotal events and partnerships, especially with big tech and stakeholders within this domain.

## 5.1 Global Majority-led AI Governance Efforts

### 5.1.1. Global AI Summit on Africa

The Global AI Summit on Africa, one of the first high-level summits within the African continent, represents a welcome shift in the trend of prominent AI convenings hosted in European countries. At the annual meeting of the World Economic Forum (WEF) in January 2024, Rwandan Minister of ICT and Innovation Paula Ingabire announced that the Rwanda Centre for the Fourth Industrial Revolution (C4IR), in conjunction with WEF will host the inaugural African high-level summit on AI to towards AI inclusivity and addressing digital divide which was initially scheduled for October 2024. (Benamara, 2024; Nzabonimpa, 2024). After rescheduling, the summit took place in April 2025 with the theme "AI and Africa's Demographic Dividend: Reimagining Economic Opportunities for Africa's Workforce" (C4IR, 2024), convening over 2,000 delegates from almost 100 countries. The summit centered on seven focus areas for discussion: people, infrastructure, data, AI models, AI applications, entrepreneurship, and governance, which traverse various sectors (C4IR, 2024). Notably, the Summit produced the Africa Declaration on Artificial Intelligence, a landmark agreement endorsed by 49 out of 55 African Union member states, the African Union secretariat, and Smart Africa (C4IR, 2025). It aims to foster ethical, inclusive, and sustainable AI development across the continent while positioning Africa as a global leader in AI governance. This declaration is supported by a proposed $60 billion Africa AI Fund, intended to de-risk early-stage ventures and build essential AI infrastructure.

The importance of hosting high-level AI summits in Africa, such as those organized by Rwanda C4IR and WEF, cannot be overstated. These summits serve as crucial platforms for African nations to proactively shape the AI landscape and harness its potential for inclusive growth and sustainable development. By fostering dialogue, knowledge sharing, and collaboration among policymakers, industry leaders, academia, and civil society, these summits can further facilitate the development of AI strategies and policies aligned with Africa's unique needs and priorities.

Moreover, increased coordination among African countries is essential to address common challenges and maximize the benefits of AI. Collaborative efforts can lead to the development of shared AI infrastructure, data sharing frameworks, and harmonized regulatory approaches, thereby creating an enabling environment for AI innovation and adoption across the continent. This coordinated approach can also amplify Africa's voice in global AI governance discussions, ensuring the continent's interests and perspectives are adequately represented.

### 5.1.2. Latin American and Caribbean Ministerial and High-Level AI Summit

In October 2023, Chile hosted the Ministerial and High Authorities Summit on the Ethics of Artificial Intelligence (AI) in Latin America and the Caribbean, which led to the Santiago Declaration on AI Ethics (Santiago Declaration, 2023; Maffioli, 2023).[8] The Summit also led to the creation of a Chile-led working group with a proposition to create an Intergovernmental Council on Artificial Intelligence for Latin America (Santiago Declaration, 2023; Maffioli, 2023). The Intergovernmental Council on Artificial Intelligence for Latin America, a first-of-its-kind global body, aims to strengthen regional AI capabilities, harmonize national AI policies, and promote the governance of AI in alignment with ethical norms. The second Ministerial and High Authorities Summit on the Ethics of Artificial Intelligence in Latin America and the Caribbean took place in Montevideo, Uruguay, from October 3-4, 2024, focusing on strengthening regional cooperation and promoting policies aligned with UNESCO's Recommendation on the Ethics of Artificial Intelligence.

Cooperation in AI within Latin America, as exemplified by the initiatives stemming from the Santiago Declaration, allows for the pooling of resources and expertise, which can be particularly beneficial for smaller countries in the region with limited AI capabilities. By working together, countries can avoid duplication of effort and leverage their collective strengths to accelerate AI development. This cooperation can also create a more consistent and predictable regulatory environment, helping address ethical concerns related to AI in a more coordinated and effective manner. Overall, a regional approach to AI governance can strengthen Latin America's position in the global AI landscape, advancing local AI ecosystems throughout the region by promoting responsible innovation that meets community needs and advances development priorities.

---

[8] See section 3.3 above to read more about the Santiago Declaration.

### 5.1.3. Commonwealth Artificial Intelligence Consortium (CAIC)

The Commonwealth of Nations is a discretionary multi-governmental organization for the advancement of democracy, enhancing economic and sustainable development, promoting peace, human rights, and the rule of law among member states. It comprises 56 countries across Africa, Asia, the Caribbean, the Americas, Europe, and the Pacific, with 87.5% of the countries from the Global Majority (The Commonwealth, 2024). In April 2023, the Commonwealth Artificial Intelligence Consortium (CAIC) was established to foster the growth and development of AI-powered innovation within the Commonwealth to address digital inequality, inclusivity, safe internet, and enhance economic and sustainable development and human rights (CAIC, 2024). The Consortium is led by Rwanda and championed by Bangladesh, the Bahamas, The Gambia, Antigua and Barbuda, Trinidad and Tobago, Mauritius, Malta, Seychelles, Sri Lanka, Fiji, and Togo (CAIC, 2024). All these countries are part of the Global Majority except Malta. The Consortium AI Ecosystem is divided into four working groups to address research and innovation, capacity building, data and infrastructure, and policy development. CAIC initiatives include the Commonwealth AI Academy, Commonwealth AI incubator, AI entrepreneurship Make-a-Thon, and Simplilearn E-learning to boost AI talents, skills, and innovation growth (CAIC, 2024). Given that these countries are united by their respective experiences under colonial subjugation, CAIC can help promote South-South cooperation amongst Global Majority countries. This cooperation, particularly within the context of AI, is essential for fostering knowledge, technology, and resource exchange to address common challenges and promote sustainable development.

# 6. Recommendations for Increasing Global Majority Inclusion in Global AI Governance & Development

Global Majority countries should be empowered to develop AI governance frameworks that reflect their specific needs and values, along with receiving opportunities to participate equitably in multilateral efforts to advance AI regulation. Since 2023, governments in the Global Majority have made significant progress towards setting frameworks and initiatives to steward AI governance, including through the African Union Continental AI Strategy, the Commonwealth Artificial Intelligence Consortium, and the Association of Southeast Asian Nations (ASEAN) Guide on AI Governance and Ethics. However, there is much more work Global Majority governments can engage in to strengthen capacity for AI governance. Potential efforts could involve building in-country policy expertise through training programs for government officials, strengthening regional AI policy networks, and establishing mechanisms for sharing best practices among Global Majority nations. A number of promising efforts have begun to ensure Global Majority representation in AI governance discourse, and this chapter aims to contribute towards the development of AI governance structures that truly reflect global perspectives and needs.

## 6.1 AI Governance Capacity Development

### 6.1.1 Developing Capacity-Building Programs

For meaningful participation in governance discussions, Global Majority governments must invest in building policy expertise in their respective countries. These efforts should go beyond traditional training programs in order to develop a complete ecosystem of policy expertise. This would include establishing AI policy research centers at local universities, creating fellowship programs for policymakers and students interested in pursuing AI governance careers, and supporting the development of local think tanks focused on AI governance. For example, the Global Center on AI Governance[9] was launched in 2024 in South Africa and has been influential in conducting empirical policy research and contributing towards AI policy development throughout the African continent. Additionally, organizations like AI Safety Asia[10] have been active in upskilling civil servants, providing them with the fundamental knowledge needed to create context-specific policies in Asian countries.

### 6.1.2 Creating Regional AI Governance Hubs

Rather than centralizing AI governance discussions in traditional centers of economic power, like London, New York, Paris, and Tokyo, establishing strong regional hubs in different Global Majority countries can enable local organizations to build technical and policy expertise while maintaining strong international connections. These hubs could be located in cities already identified as emerging AI talent centers like Nairobi, São Paulo, or Jakarta, as highlighted in Section 2. Along with advancing local AI development, these hubs would serve as gathering points for refining regional expertise, advancing policy development, and enabling the coordination of broader initiatives. Activities conducted within these hubs would also feed directly into global governance initiatives while helping policymakers and other stakeholders maintain strong connections to local contexts and needs.

---

[9] https://www.globalcenter.ai/
[10] https://www.aisafety.asia/

### 6.1.3 Establishing Financial Support Mechanisms

Meaningful participation in AI development and governance will require a substantial number of resources, including financial resources that support government, civil society, and academia stakeholders. Global Majority countries should invest in creating dedicated funding mechanisms that enable their respective stakeholders to participate fully in governance discussions. This could include travel support for convenings, funding for local research and policy development, and resources for implementing governance frameworks. Additionally, Global Majority countries should advocate for restructuring AI funding mechanisms to be more accessible while developing innovative financing models like AI development bonds. These bonds could be issued by a specialized entity, providing a stable funding stream for AI research and development within these countries.

### 6.1.4 Restructuring International Decision-Making Processes and Creating Meaningful Participation Mechanisms

Currently, most AI governance discussions happen in forums dominated by Western nations and institutions. To change this, Global Majority countries could steer efforts toward fundamentally redesigning how these discussions occur. Similar to the operating structures of the G7 and G20, these countries should establish rotating leadership positions to create regional working groups with real decision-making power and ensure that the resulting discussions address Global Majority priorities. Most importantly, countries would work together to ensure that their needs are not sidelined in favor of large companies at the forefront of AI development. Such efforts would ensure genuine participation from Global Majority countries and that priorities and perspectives from Global Majority countries regularly drive global AI governance agendas.

## 6.2 Advancing AI Policy Development

### 6.2.1 Creating Inclusive Technical Standards

Technical standards often form the backbone of governance frameworks. To ensure these standards work for everyone, Global Majority countries must play a significant role in their development, particularly as these efforts are initiated. To help facilitate such participation, countries could create working groups that specifically focus on how AI standards can work in different contexts, from areas with limited internet connectivity to regions with different approaches to data privacy. These groups could help shape work that investigates low-bandwidth data processing and storage approaches to facilitate AI adoption in resource-constrained contexts, along with investigating how AI standards can be leveraged more broadly to support economic development and social progress.

### 6.2.2 Implementing Knowledge and Resource Transfer Mechanisms

There is a strong need for systematic ways to share knowledge and resources between different parts of the global AI governance community. While knowledge transfer often happens organically, systemizing these efforts could lead to more strategic partnerships and improved policy outcomes. Global Majority countries could be involved in creating dedicated funding streams for joint research projects, establishing mentorship programs between experienced and emerging policymakers, and developing shared resources in multiple languages. These efforts would not only be siloed to government entities but should also enable academic institutions and other local organizations to facilitate knowledge transfer.

### 6.2.3 Building Accountability Mechanisms

Global Majority countries need strong accountability mechanisms to ensure these inclusive approaches to global AI governance are effective. This would involve creating clear metrics for measuring inclusion in global initiatives, a regular review of formal processes, and mechanisms for Global Majority countries to raise concerns when systems do not function as intended. Establishing independent oversight bodies or committees with equitable representation from Global Majority countries can further help facilitate dialogue and collaboration to advance the goals of international AI cooperation. These mechanisms would encourage equitable participation in AI governance processes while ensuring tangible progress towards AI oversight.

## 7. Conclusion

The global governance of artificial intelligence is at a crossroads, presenting significant challenges and opportunities for reshaping power dynamics between the Global Majority and the historically dominant countries in the West. As this chapter has illustrated, the disparities in AI development and governance are deeply entrenched, rooted in systemic inequalities in education, digital infrastructure, economic resources, and access to decision-making platforms. These disparities perpetuate a technological and geopolitical dependency that undermines the agency of Global Majority countries and risks deepening existing global divides. However, the emergence of regional governance frameworks, grassroots initiatives, and multilateral collaborations signals a growing recognition of the need for more inclusive approaches. Countries in the Global Majority and regional bodies like the African Union and ASEAN are actively working to define their governance paradigms, ensuring that AI technologies align with their unique cultural, economic, and ethical priorities. These efforts, while promising, require sustained support, including investment in capacity building, resource redistribution, and the restructuring of global governance forums to amplify diverse perspectives.

For AI governance to truly serve as a force for global equity, it must transcend superficial inclusion and embrace systemic reforms. This entails integrating Global Majority priorities into international standards, fostering meaningful participation in decision-making processes, and addressing the structural barriers that limit access to the benefits of AI. By building mechanisms for accountability, resource-sharing, and equitable representation, the global community can work toward a future where AI governance reflects the needs and aspirations of all humanity, not just a privileged few. Ultimately, the success of global AI governance depends on its ability to balance innovation with inclusivity, safety with sovereignty, and efficiency with equity. This chapter underscores the imperative for collaborative action to ensure that AI does not reinforce historical patterns of exclusion but instead becomes a tool for collective progress and shared prosperity.

Okolo, C. T. (2024a). Operationalizing Comprehensive Data Governance in Africa. *African Journal of Sustainable Development*, *14*(1), 200.

Okolo, C. T. (2024b). *Reforming data regulation to advance AI governance in Africa*. Brookings Institution. https://www.brookings.edu/articles/reforming-data-regulation-to-advance-ai-governance-in-africa/

Organisation for Economic Co-operation and Development. (2019). *Recommendation of the Council on Artificial Intelligence*. OECD. https://legalinstruments.oecd.org/en/instruments/OECD-LEGAL-0449

Pacific Ministers for Information and Communications Technologies. (2023). *Pacific ICT Ministerial Declaration: Lagatoi Declaration on the Digital Transformation of the Pacific.* Pacific ICT Ministerial Dialogue. https://www.ict.gov.pg/Press%20Statement/Pacific%20ICT%20Ministerial%20Declaration.%20Monday%2028%20August%202023.%20APEC%20Haus.pdf

Peng, L. (2024). *Artificial Divides: Global AI Access Disparities and Constructions of New Digital Realities* (Master's thesis, University of Washington).

Prabhakaran, V., Qadri, R., & Hutchinson, B. (2022). Cultural incongruencies in artificial intelligence. arXiv preprint arXiv:2211.13069.

Putra, B. A. (2024). Governing AI in Southeast Asia: ASEAN's way forward. *Frontiers in Artificial Intelligence*, *7*. https://doi.org/10.3389/frai.2024.1411838

QS World University Rankings. (2024). *QS World University Rankings Data Science and Artificial Intelligence.* QS Quacquarelli Symonds. https://www.topuniversities.com/university-subject-rankings/data-science-artificial-intelligence
31

33